\documentclass{aa}

\usepackage{graphicx}
\usepackage{natbib}
\usepackage{amsmath}
\bibpunct{(}{)}{;}{a}{}{,}
\usepackage{txfonts}

\begin{document}

\title{Magnetic field generation in relativistic shocks}
\subtitle{%
An early end of the exponential Weibel instability in
electron-proton plasmas}
\author{J.~Wiersma
    \and A.~Achterberg}
\institute{Sterrenkundig Instituut, Universiteit Utrecht, P.O.~Box 80000,
    NL-3508 TA Utrecht, The Netherlands}
\offprints{J.~Wiersma,\\
    \email{wiersma@astro.uu.nl}}
\date{Received / Accepted }

\abstract{%
We discuss magnetic field generation by the proton Weibel
instability in relativistic shocks, a situation that applies to the
external shocks in the fireball model for Gamma-ray Bursts, and possibly
also to internal shocks.
Our analytical estimates show that the linear phase of the instability ends
well before it has converted a significant fraction of the energy in the
proton beam into magnetic energy: the conversion efficiency is much smaller
(of order $m_{\rm e}/m_{\rm p}$) in electron-proton plasmas than in pair
plasmas.
We find this estimate by modelling the plasma in the shock transition zone
with a waterbag momentum distribution for the protons and with a background
of hot electrons.

For ultra-relativistic shocks we find that the wavelength of the most
efficient mode for magnetic field generation equals the electron skin depth,
that the relevant nonlinear stabilization mechanism is magnetic trapping,
and that the presence of the hot electrons limits the typical magnetic field strength generated by this mode so that it does not
depend on the energy content of the protons.
We conclude that other processes than the linear Weibel instability must
convert the free energy of the protons into magnetic fields.
}

\maketitle

\section{\label{sec:intro}Introduction}

Magnetic field generation in relativistic shocks in a hydrogen
(electron-proton) plasma is important for the fireball model for Gamma-ray
Bursts~\citep{reme92}.
This model proposes that the non-thermal radiation we observe in the prompt
and afterglow emission from the Gamma-ray Burst is synchrotron radiation
from collisionless relativistic shocks.
To explain the observed intensity of the afterglows as synchrotron emission,
the models need a magnetic field strength~$B$ of at least ten per cent of the equipartition
field strength~\citep{grwa99,paku02}.
This means that the magnetic energy density must contribute about one to
ten per cent to the total energy density of the plasma behind the shock:
$B^{2}/8 \pi\sim(0.01-1) \times e$, with $e$ the post-shock thermal
energy density.
In electron-proton plasmas this implies a much stronger magnetic field than
in pair plasmas because the energy density in an ultra-relativistic shock
propagating into a cold medium (where $n m c^{2} \gg P$, with $n$ the number
density, $m$ the rest mass and $P$ the thermal pressure)
is roughly proportional to the rest mass $m$ of the particles.
In this paper, we look at what happens in a plasma consisting of particles of widely different mass.

In collisionless shocks, plasma instabilities can generate magnetic fields.
Within the shock transition layer the relative motion of the mixing
pre- and post-shock plasma produces very anisotropic velocity
distributions for all particle species concerned.
Fluctuating electromagnetic fields deflect the incoming charged particles
and act as the effective collisional process needed to complete the shock
transition~\citep[{\it e.g.},][]{hagl92}.
These fluctuating fields occur naturally because anisotropic velocity
distributions are unstable against several plasma instabilities, such as
the electrostatic two-stream instability and the electromagnetic Weibel
instability.
The first is an instability of (quasi-)longitudinal charge density
perturbations and leads to fluctuating electric fields satisfying $|\mbox{\boldmath$ E $} |
\gg|\mbox{\boldmath$ B $} |$ where $\mbox{\boldmath$ E $}$ and $\mbox{\boldmath$ B $}$ are the fluctuating electric and
magnetic fields.
The second is an instability of the advection currents (proportional to the
beam velocity) that result from charge bunching in the beams, and leads to
spontaneously growing transverse waves~\citep{weib59} with $|\mbox{\boldmath$ B $}| \ge|\mbox{\boldmath$ E $}|$.
In a relativistic shock, where the relative velocity of the pre- and
post-shock plasma approaches the velocity of light, the Weibel instability
dominates because it has the largest growth rate \citep{ccc02}.

As both analytical estimates and numerical simulations show, the Weibel
instability in pair plasmas can produce a magnetic field of
near-equipartition strength \citep{fstm03,hasa03,ksnb98,melo99,yal94}.
In numerical simulations, the magnetic field generation always undergoes an
exponentially growing phase that agrees with the estimates from linear
analytical theory, and enters a nonlinear phase after that.
\citet{yal94} have shown that in pair plasmas, the magnetic field
strength reaches its maximum value at the end of the linear phase of the
instability.
The question arises whether the same holds true for the Weibel instability
operating in an electron-proton plasma~\citep{melo99}.
Numerical simulations \citep{fhhn04} show that the
nonlinear phase may be more important in electron-ion plasmas.

In this paper we present an analytical estimate that shows that the
linear phase of the instability ends much earlier for proton beams
in a hydrogen plasma than for electron(-positron) beams in a pair plasma.
We do not present a full self-consistent shock model: rather we consider
the plasma processes that could generate a magnetic field in a plasma with
properties such as one expects near the front of a collisionless
relativistic shock.
The paper is organised as follows.
In Sect.~\ref{sec:model} we define a model for the shock situation in terms
of the momentum distributions of the particles.
We calculate the conditions for the instability in Sect.~\ref{sec:growthrate} and
the magnetic field strength at the end of the linear phase of the
instability in Sect.~\ref{sec:stab}.
In Sect.~\ref{sec:epsB}, we compare the energy density associated with this
magnetic field strength with the total available energy density associated
with the beams.
Section~\ref{sec:discussion} contains the discussion and Sect.~\ref{sec:conclusions}
the conclusions.

\section{\label{sec:model}A simple model for a relativistic shock transition}

In an electron-proton plasma, the protons dominate the shock energetics
because they have a much larger rest mass than the electrons.
Therefore, we will study the proton-driven Weibel instability.
In this section we present a simple model for the plasma in the
transition layer at the front of the shock.

The plasma in an astrophysical relativistic shock does not necessarily
behave as a {\em single} fluid.
Coulomb collisions between electrons and protons are not sufficiently fast
to create thermal equilibrium between the protons and electrons.
This problem of the non-equilibration of the electron and ion energies
already exists in the much slower ($\sim 1000$ km/s) shocks associated with
Supernova Remnants~\citep{drmc93,vink04}.

We assume that scattering by plasma waves is far more
efficient for the light electrons than for the heavy ions so that
when the trajectories of the incoming protons start to become significantly
perturbed, the electrons have already undergone the fast-growing
electron Weibel instability~\citep{fhhn04,melo99},
which has converted the kinetic energy of their bulk motion into the thermal
energy of a relativistically hot electron plasma with an (almost) isotropic
thermal velocity distribution.
The incoming protons form, seen from the rest frame of the hot electrons, a
relativistic beam.
We also assume that part of the protons are reflected further downstream
although that assumption is not critical for our final
conclusions (see Sect.~\ref{sec:asym}).

The electron-driven Weibel instability produces a weak fluctuating magnetic
field with $B^{2}/8 \pi\sim e_{\rm e}$, with $e_{\rm e}$ the energy density
of the shocked electrons.
We ignore this magnetic field in the calculations for proton beams, but it
could serve as a seed perturbation for the proton-driven Weibel instability.

\subsection{\label{sec:fp}The proton velocity distribution}

A simple model for the anisotropic proton velocity distribution within the shock transition
layer is a {\em waterbag distribution} \citep{sftm02,yoda87}.
We consider a similar situation as in Fig.~6 of~\citet{fhhn04}: we take two
counter-streaming proton beams moving along the $x$-direction, with a small
velocity spread in the $z$-direction to model thermal motions:
\begin{equation}
\label{eq:distribution}
\begin{split}
 F(\vec{p}) ={} &{\frac{n_{\rm p}}{4 p_{z0}}} \:
	\left[ \delta(p_x-p_{x0}) + \delta(p_x+ p_{x0}) \right]\\
	&{}\times\delta(p_y) \left[\Theta(p_z+p_{z0}) - \Theta(p_z-p_{z0}) \right]\,.
\end{split}
\end{equation}
Here $n_{\rm p}$ is the total proton density, $p_{x0}$ is the bulk
momentum of the proton beams, $p_{z0}$ is the maximum momentum in thermal
motions and $\Theta(x) = (1 + x/|x|)/2$ is the Heavyside step function.
The assumption of two beams of equal strength is mathematically convenient,
but not essential (see Sect.~\ref{sec:asym}).

This is a simple model that mimics the properties of non-relativistic
collisionless shocks \citep[see the {\em Microstructure} Section
in][]{tsst85} in which (partial) reflection of the ions occurs as a
result of deflection by an electrostatic potential jump in the shock
transition, or by `overshoots' in the strong magnetic field in the wake of
the shock.
In addition, the waterbag model accounts for partial ion heating by
including a velocity dispersion in the direction perpendicular to both the
beam direction and the wave magnetic field.
This direction lies along the wave vector of the unstable modes (the
$z$-direction in our configuration).

\subsection{\label{sec:shockconditions}The shock conditions for the electrons}

We assume that the electrons have (almost) completed the shock transition so
that their properties obey the relativistic shock conditions \citep{blmc76},
which follow from the generally valid conservation laws for particle
number, energy and momentum.

Here and below, we will label properties of the post-shock electron plasma
with subscript~$2$, and those of the pre-shock plasma with subscript~$1$.
We will assume that the pre-shock plasma is cold in the sense that
$e_{{\rm e},1} \ll n_{{\rm e},1} m_{\rm e} c^{2}$.
Then the shock conditions for the proper density~$n_{\rm e}$ and the
proper energy density~$e_{\rm e}$ for the electrons are:
\begin{equation}
\label{eq:shockn}
\begin{aligned}
 n_{{\rm e},2}	&= (4 \gamma _{\rm rel}+3 )n_{{\rm e},1},		\\
 e_{{\rm e},2}	&= (4 \gamma _{\rm rel}+3 )\gamma _{\rm rel}n_{{\rm e},1} m_{\rm e} c^2,
\end{aligned}
\end{equation}
where $\gamma _{\rm rel}^2 = 1 + u_{x0}^2$ is the Lorentz factor of the relative
velocity between the pre- and post-shock plasma (with $u_{x0}=p_{x0}/m_{\rm p}c$).

We neglect the dynamical influence of a pre-shock magnetic field on the
electron-fluid jump conditions.
This influence will be small if $V_{\rm Ae, 1} \ll c$ where $V_{\rm Ae, 1}
\equiv B_{1}/\sqrt{4 \pi n_{e,1} m_{\rm e}}$ is the Alfv\'en speed based on
the electron mass (instead of the proton mass).
We also neglect any large-scale electrostatic field in the shock that might
accelerate the electrons to higher energies while decelerating the
incoming protons.

\section{\label{sec:weibel}The Weibel instability}

\subsection{\label{sec:growthrate}The dispersion relation}

We consider the instability of a purely transverse electromagnetic wave with
wave vector $\vec{k}= k \vec{e}_z$ and frequency~$\omega$.
The instability will grow with a rate equal to the imaginary part of the
wave frequency:~$\sigma= \Im(\omega)$.

The Weibel instability for the model of the previous section obeys a
dispersion relation of the form
\citep[{\it e.g.},][]{sftm02}
\begin{equation}
\label{eq:symmdisp}
 k^2 c^2 - \omega^{2} \left[ \: 1 + \chi _{xx}(\omega, k) \: \right] = 0\,,
\end{equation}
where $\chi _{xx}(\omega, k) \equiv\chi _{xx,{\rm e}}(\omega, k) + \chi _{xx,{\rm p}}(\omega, k)$ is the $xx$-component of the plasma susceptibility
tensor, which contains contributions of both the electrons and the protons.

Since the electrons have a relativistically hot thermal velocity
distribution their contribution is
\begin{equation}
 \chi _{xx,{\rm e}}(\omega, k) = - \frac{\tilde{\omega}_{\rm pe}^2}{\omega^{2}}\,,
 \label{eq:cxxze}
\end{equation}
where $\tilde{\omega}_{\rm pe}$ is the electron plasma frequency (in Gaussian units):
\begin{equation}
	\tilde{\omega}_{\rm pe}^{2} = \frac{4\pi q^2 n_{{\rm e},2}}{m_{{\rm e}}h}\,,
\end{equation}
with $n_{{\rm e}}$ the electron proper density, $q$ the electron charge, $m_{{\rm e}}$ the
electron mass and $h= (e_{\rm e} + P_{\rm e})/(n_{\rm e} m_{{\rm e}}c^{2})
\simeq 4e_{\rm e}/(3 n_{\rm e} m_{{\rm e}}c^{2})$ the electron enthalpy per
unit rest energy for the relativistically hot electrons with $e_{\rm e}
\simeq 3 P_{\rm e} \gg n_{\rm e} m_{\rm e} c^2$.
If we assume that the electrons are fully shocked, the shock conditions
(\ref{eq:shockn}) enable us to express $\tilde{\omega}_{\rm pe}$ in terms of the pre-shock
electron number density:
\begin{equation}
\label{eq:omper}
 \tilde{\omega}_{\rm pe}^2 = \frac{12\pi q^2 n_{{\rm e},1}}{m_{{\rm e}}} \:
	\left({\frac{4\gamma _{\rm rel}+ 3}{4\gamma _{\rm rel}}} \right)\,.
\end{equation}
The factor between brackets approaches unity in ultra-relativistic
shocks with $\gamma _{\rm rel}\gg 1$.

The proton contribution to dispersion relation~(\ref{eq:symmdisp}) is
\citep[see][]{sftm02}
\begin{equation}
 \chi _{xx,{\rm p}}(\omega, k) = - \frac{\omega _{\rm pp}^2}{\gamma _{\rm b0}\: \omega^{2}} \: \left(
	{\cal F}+ \frac{k^2 v_{x0}^2}{\omega^2 - k^2 v_{z0}^2} \right),
 \label{eq:cxxzp}
\end{equation}
with
\begin{equation}
 {\cal F}= \frac{c}{2 v_{z0}} \: \ln\left(\frac{c+v_{z0}}{c-v_{z0}} \right)
	- \frac{u_{x0}^2}{1+ u_{x0}^2}.
 \label{eq:F}
\end{equation}
Here $\omega _{\rm pp}= \sqrt{4\pi q^2n_{\rm p}/m_{\rm p}}$ is the (non-relativistic) proton
plasma frequency based on the density in the lab frame, $m_{\rm p}$ is the proton
rest mass, $u_i = p_i/(m_{\rm p}c)$, $\gamma _{\rm b0}= (1 + u_{x0}^2 + u_{z0}^2)^{1/2}$ and
$v_i = u_i c/\gamma _{\rm b0}$.
To ensure quasi-neutrality of the plasma we must have
\begin{equation}
\label{eq:quasin}
	n_{\rm p} \approx n_{{\rm e},2}\,,
\end{equation}
and the associated plasma frequency is
\begin{equation}
\label{eq:pfreq1}
 \omega _{\rm pp}^2 = (4 \gamma _{\rm rel}+ 3) \: \frac{4\pi q^2 n_{\rm e,1}}{m_{\rm p}}\,.
\end{equation}

For what follows it is convenient to introduce the frequency $\hat{\omega}_{\rm pp}$
defined by
\begin{equation}
\label{eq:pfreq2}
	\hat{\omega}_{\rm pp}^{2} = \frac{\omega _{\rm pp}^2}{\gamma _{\rm b0}}\,.
\end{equation}
If the velocity dispersion in the beam is small, which is always true
for a Weibel-unstable proton distribution (see the end of this section), we
have $\gamma _{\rm b0}\simeq\gamma _{\rm rel}$ so that
\begin{equation}
\label{eq:plfreqratio}
	\frac{\hat{\omega}_{\rm pp}^{2}}{\tilde{\omega}_{\rm pe}^2} \simeq\frac{4 m_{\rm e}}{3 m_{\rm p}}\,.
\end{equation}

Note that one can get the equations for an electron-positron beam in an
electron-positron plasma \citep[see also][]{sftm02,yal94} by replacing $n_{\rm p}$
and~$m_{\rm p}$ with the beam density and electron mass respectively.
We will use this in what follows to compare results for electron-proton
plasmas with those for electron-positron plasmas.
In those cases we assume that the density of the electron-positron beams is
comparable to the density of the background electron-positron plasma.

\begin{figure}
\resizebox{\hsize}{!}{\includegraphics{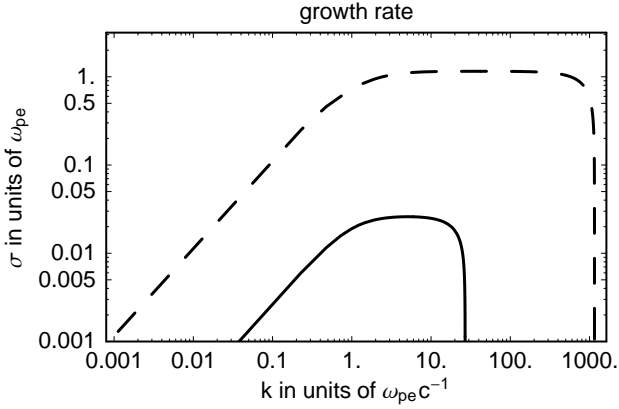}}
\caption{Growth rate as a function of wave number for
a shock with $\gamma _{\rm b0}= 1000$ and $v_{z0}= 0.001 c$.
The solid line is for proton beams in a background of hot electrons; the
dashed line is for electron-positron beams in an electron-positron plasma.}
\label{fig:sigma} \end{figure}

Substituting the contributions (\ref{eq:cxxze}) and~(\ref{eq:cxxzp}) in~(\ref{eq:symmdisp}) we
can write the dispersion relation as a biquadratic equation for~$\omega$:
\begin{equation}
 \omega^4 - {\cal B}\omega^2 + {\cal C}= 0\,,
 \label{eq:biquadratic}
\end{equation}
with
\begin{align}
 {\cal B}&= k^2 (c^2 + v_{z0}^2) + \Omega^2,	\\
 & \nonumber\\
 {\cal C}&= k^2\{v _{z0}^2(k^2 c^2 + \Omega^2) - \hat{\omega}_{\rm pp}^2v_{x0}^2\},
\end{align}
where
\begin{equation}
	\Omega^2 \equiv\tilde{\omega}_{\rm pe}^2 + \hat{\omega}_{\rm pp}^2 \: {\cal F}.
	\label{eq:Om}
\end{equation}
Since ${\cal B}> 0$ the wave is unstable for ${\cal C}< 0$ with $\omega^{2} \equiv-
\sigma^{2} < 0$ where
the growth rate $\sigma$ follows from
\begin{equation}
\label{eq:sigma}
 \sigma^2 = \frac{\displaystyle\sqrt{{\cal B}^2 - 4 {\cal C}} - {\cal B}}{2}\,.
\end{equation}
For a given set of shock parameters ($\gamma _{\rm b0}$, $v_{z0}$, $n_0$), the growth
rate is a function of the wave number (Fig.~\ref{fig:sigma}).

Anticipating our results for a proton beam in a background of
(relativistically) hot electrons, we will assume that the growth rate of the
unstable modes satisfies
\begin{equation}
\label{eq:smallcond}
	\sigma\ll kc \quad\mbox{and} \quad\sigma\ll\tilde{\omega}_{\rm pe}\,,
\end{equation}
and that the characteristic plasma frequencies satisfy
\begin{equation}
\label{eq:plfreqcond}
	\hat{\omega}_{\rm pp}^2 \ll\Omega^{2},
\end{equation}
see Eq.~(\ref{eq:Omapprox}) below.

Under these assumptions we can approximate the solution of the dispersion relation with
$\sigma^{2} \simeq- {\cal C}/{\cal B}$, which leaves the instability criterion (${\cal C}<
0$) unchanged.
We will also make the approximation of a small beam velocity
dispersion: $v_{z0}^{2} \ll c^{2}$. Then the dispersion relation
reduces to
\begin{equation}
\label{eq:approxdisp2}
	\sigma^{2} = {k^2(\hat{\omega}_{\rm pp}^2 \: v_{x0}^2 - \Omega^2 \: v_{z0}^2 - k^2 c^{2} \: v_{z0}^2)
		\over k^2 c^2 + \Omega^2}.
\end{equation}

For further analysis we introduce the following dimensionless quantities:
\begin{equation}
\label{eq:dimlvar}
	\kappa= \frac{kc}{\Omega}\,, \quad\nu= \frac{\sigma}{\Omega}
\,, \quad\alpha= \frac{\hat{\omega}_{\rm pp}}{\Omega} \: \left( \frac{v_{x0}}{v_{z0}} \right)\,,
\end{equation}
Expressed in these quantities the dispersion relation~(\ref{eq:approxdisp2}) reads
\begin{equation}
\label{eq:approxdisp3}
 \begin{aligned}
	\nu^{2} & = \left( \frac{v_{z0}}{c} \right)^{2}
	\frac{\displaystyle\kappa^{2} \: \left( \kappa _{\rm max}^{2} - \kappa^{2} \right) }{1 + \kappa^{2}}
	\\
	& \\
	& = \left(\frac{\hat{\omega}_{\rm pp}}{\Omega} \right)^{2} \:
	\left( \frac{v_{x0}}{c} \right)^{2} \:
	\frac{\kappa^{2}(\kappa _{\rm max}^{2} - \kappa^{2})}
	{(1 + \kappa^{2})(1 + \kappa _{\rm max}^{2})}
	\,,
 \end{aligned}
\end{equation}
where we have eliminated $v_{z0}$ using the definition
of~$\alpha$ and we define
\begin{equation}
\label{eq:kmax}
	\kappa _{\rm max} \equiv\sqrt{\alpha^{2} - 1},
\end{equation}
which is the limiting wavenumber of the Weibel instability: the instability
condition is satisfied for $\kappa< \kappa _{\rm max}$.
The parameter $\alpha$ is a measure for the range of unstable wave numbers.
The Weibel instability occurs only if $\alpha> 1$.
This condition poses a restriction on the velocity spread~$v_{z0}$:
\begin{equation}
 {v_{z0}\over v_{x0}} < {\hat{\omega}_{\rm pp}\over\Omega}
 \label{eq:smallvzn1}
\end{equation}
The quantity $\Omega$ depends on $v_{z0}$ through Eqs. (\ref{eq:Om}) and~(\ref{eq:F}), but
a quick inspection shows that ${\cal F}$ and $\Omega$ are
increasing functions of $v_{z0}$ so if condition~(\ref{eq:smallvzn1}) holds,
then we also have
\begin{equation}
 {v_{z0}\over v_{x0}} < {\hat{\omega}_{\rm pp}\over\Omega|_{v_{z0}=0}}
 = {\hat{\omega}_{\rm pp}\over(\tilde{\omega}_{\rm pe}^2 + \hat{\omega}_{\rm pp}^2/\gamma _{\rm rel}^2)^{1/2}}\,.
 \label{eq:smallvzn2}
\end{equation}
It follows that for relativistic shocks with $v_{x0}\to c$ the proton-driven
instability requires $v_{z0}\ll c$ so that
\begin{equation}
	{\cal F}\simeq\frac{1}{\gamma _{\rm b0}^{2}} + \frac{v_{z0}^{2}}{3c^{2}} \ll 1\,,
\end{equation}
and
\begin{equation}
 \Omega\simeq\tilde{\omega}_{\rm pe}\gg\hat{\omega}_{\rm pp}\label{eq:Omapprox}
\end{equation}
(see Eq.~\ref{eq:plfreqratio}).

The largest growth rate occurs for a mode with wave number
$k_{\ast}$, which
follows from $({\rm d} \sigma/{\rm d} k)_{k=k_{\ast}} = 0$.
Using the approximated dispersion relation~(\ref{eq:approxdisp3}) we find that
this wavenumber equals
\begin{equation}
\label{eq:kast}
	\frac{k_{\ast}c}{\Omega} \equiv\kappa _{\ast}(\alpha) =
	\sqrt{\alpha- 1}\,,
\end{equation}
with the corresponding growth rate
\begin{equation}
\label{eq:nuast}
 \begin{aligned}
	\frac{\sigma(k_\ast)}{\Omega} & \equiv\nu _{\ast} = (\alpha-1) \: \frac{v_{z0}}{c} \\
	& =
	\frac{\alpha-1}{\alpha} \: \left(\frac{\hat{\omega}_{\rm pp}}{\Omega}
	\right) \:
	\frac{v_{x0}}{c}\,.
 \end{aligned}
\end{equation}
In the last equality we have used the definition of $\alpha$.
For a strong, relativistic shock we find $\sigma(k_\ast)\simeq\hat{\omega}_{\rm pp}$.

In view of this fact and Eq.~(\ref{eq:Omapprox}), conditions~(\ref{eq:smallcond})
and~(\ref{eq:plfreqcond}) automatically hold for the proton-driven Weibel
instability.

\subsection{\label{sec:stab}Stabilization of the Weibel instability}

The linear phase of the Weibel instability (during which perturbations grow
exponentially with time) ends when the generated electromagnetic fields
significantly perturb the trajectories of the particles taking part in the
instability.
Because the magnetic fields generated by the instability are inhomogeneous,
the particles will quiver under the influence of the Lorentz force.
When the amplitude of these quiver motions exceeds the wavelength of the
instability, the linear theory breaks down.
\citet{yal94} give a full treatment of these quiver motions and show that
this criterion agrees with the magnetic trapping argument, which says that
the instability will stop when the wave magnetic field becomes so strong
that it traps the beam particles.

\begin{figure}
\resizebox{\hsize}{!}{\includegraphics{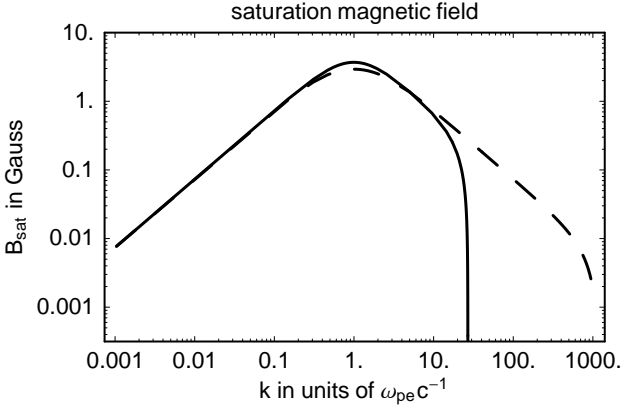}}
\caption{Saturation magnetic field as a function of wave number
for a proton-driven instability and for an
instability in an electron-positron plasma, using the same parameters as in
Fig.~\ref{fig:sigma} with $n_{e,1} = 1\,\hbox{cm}^{-3}$.}
\label{fig:bsat} \end{figure}

The linearized equation describing the quiver motions in the $z$-direction
for a beam particle in a magnetic field $B(z,t) \: \mbox{\boldmath$ e $}_{y}$ reads:
\begin{equation}
	\frac{{\rm d}^{2} \xi _{z}}{{\rm d} t^{2}} =
	\frac{q v_{x0} \: B(z,t)}{\gamma _{\rm b0} m_{\rm p} c}\,,
\end{equation}
where $\xi _{z}$ is the displacement.
In the linear stage of the instability the wave magnetic field varies as
$B(z,t) = |\mbox{\boldmath$ B $}|\exp(\sigma t)\sin(k z)$ for a wave with wave number $k$ and
growth rate $\sigma$, so the amplitude of the quiver motion is
\begin{equation}
\label{eq:quiverampl}
	| \xi _{z} | \sim\frac{q |\mbox{\boldmath$ B $}|v_{x0}}{\gamma _{\rm b0} m_{\rm p} c \sigma^{2}}\,.
\end{equation}
The trapping criterion $k |\xi _{z}| < 1$ corresponds to $|\mbox{\boldmath$ B $}|<
B_{\rm trap}$ with
\begin{equation}
 \label{eq:trapcond}
 B_{\rm trap}= \frac{\gamma _{\rm b0}m_{\rm p}c\sigma^{2}(k)}{kv_{x0}q}\,.
\end{equation}
This corresponds to Eq.~(18) of~\citet{yal94}.
Assuming that trapping saturates the Weibel instability at {\em all}
wavelengths we get the typical field amplitude as
a function of wavenumber (Fig.~\ref{fig:bsat}).

The maximum field amplitude is reached at those wave numbers where $\sigma^{2}/k$ has the maximum value.
For dispersion relation~(\ref{eq:approxdisp3}) this maximum is reached at a
wavenumber $k^\dagger$ that follows from
\begin{equation}
\label{eq:kbar}
	\frac{k^\dagger c}{\Omega} \equiv\kappa^\dagger(\alpha) = \left\{
	\frac{\alpha}{2} \: \sqrt{8 + \alpha^{2}} - 1 - \frac{\alpha^{2}}{2}
	\right\}^{1/2}
\,.
\end{equation}
If the instability is strong, so that $\alpha\gg 1$, we find that this wavenumber and the
associated growth rate are
\begin{equation}
\label{eq:strong2}
	\frac{k^{\dagger} c}{\Omega} \simeq 1\,\; ,\,\;
	\frac{\sigma(k^{\dagger})}{\Omega} \simeq\frac{\alpha}{\sqrt{2}} \: \frac{v_{z0}}{c}
	=\left(\frac{\hat{\omega}_{\rm pp}}{\sqrt{2} \: \Omega}
	\right) \:
	\frac{v_{x0}}{c}\,.
\end{equation}
Note that the trapping criterion predicts the largest field amplitude at a
wavenumber $k^{\dagger}$, which is not equal to the wavenumber $k_{\ast}$ of
the fastest growing mode (Eq.~\ref{eq:kast}): $k^{\dagger} \simeq\Omega/c
\ll k_{\ast} \simeq\sqrt{\alpha} \Omega/c$ for $\alpha\gg 1$.
Eventually, this slower growing long-wavelength mode reaches a higher
magnetic field than the faster growing mode
\citep[see also the discussion in][]{yal94}.
If $\alpha\gg 1$ the saturation field strength due to trapping at
$k^{\dagger} \simeq\Omega/c$ is
\begin{equation}
\label{eq:trapsat}
\begin{aligned}
	B_{\rm trap}&\simeq\frac{\gamma _{\rm b0}m_{\rm p}v_{x0}}{q}\:\frac{\hat{\omega}_{\rm pp}^2}{2\Omega}\\
	 &\simeq{2\gamma _{\rm b0}m_{{\rm e}}\tilde{\omega}_{\rm pe}c\over 3q},
\end{aligned}
\end{equation}
where we have used Eqs.~(\ref{eq:plfreqratio}) and~(\ref{eq:Omapprox}) to eliminate
$\hat{\omega}_{\rm pp}$ and~$\Omega$.

For parameters typical for the external shock associated with Gamma-ray
Bursts we have
\begin{equation}
\label{eq:bsat-ur}
 B_{\rm trap}\simeq 3.7 \sqrt{n_1} \: \gamma _{1000}\, \ifmmode{\rm Gauss}\else${\rm Gauss}$\fi,
\end{equation}
with $n_1 = n_0/ (1\,\ifmmode{\rm cm^{-3}}\else${\rm cm^{-3}}$\fi)$ and $\gamma _{1000} = \gamma _{\rm b0}/1000$.

We should note that \cite{melo99} use a different method for estimating
$B$: they propose that the instability saturates when the beam ions become
magnetized.
This happens when the Larmor radius~$r_{\rm L}= \gamma v m c^2/qB$ of the beam
particles in the generated magnetic field becomes smaller than the
wavelength of the fastest growing mode of the instability.
This criterion $k_{\ast}r_{\rm L}< 1$ corresponds to $|\mbox{\boldmath$ B $}|< B_{\rm magn}$ with
\begin{equation}
\label{eq:MedvLoe}
	B_{\rm magn}= {\gamma _{\rm b0}v_{x0}m_{\rm p}c k_\ast\over q},
\end{equation}
with $k_\ast$ given by Eq.~(\ref{eq:kast}).
However, the trapping argument predicts the smallest saturation
amplitude $|\mbox{\boldmath$ B $}|$.
In particular, we have
\begin{equation}
\label{eq:satratio}
	\frac{B_{\rm trap}}{B_{\rm magn}} =
	\frac{\sigma^{2}(k^\dagger)}{k^{\dagger} k_{\ast} v_{x0}^{2}}
	= \left( \frac{\hat{\omega}_{\rm pp}}{\Omega} \right)^{2} \: \Phi(\alpha),
\end{equation}
where
\begin{equation}
\label{eq:function}
	\Phi(\alpha) \equiv\frac{\kappa^\dagger(\alpha)} {\alpha^{2} \: \sqrt{\alpha- 1}} \:
	\left(
	\frac{3 \alpha- \sqrt{8 + \alpha^{2}}}{\sqrt{8 + \alpha^{2}} - \alpha}
	\right)\,.
\end{equation}
To derive this relation we have used definition~(\ref{eq:dimlvar}) to write
$v_{z0}/v_{x0} = \hat{\omega}_{\rm pp}/(\alpha\Omega)$.
For a proton beam in a hot electron background we have
$\hat{\omega}_{\rm pp}^{2}/\Omega^{2} \ll 1$ and $\Phi(\alpha) < 0.3$ for all $\alpha\ge 1$.
For a strong instability with $\alpha\gg 1$ we have $\Phi(\alpha) \simeq(4\alpha)^{-1/2} \ll 1$.
Therefore, trapping occurs well before the field can totally magnetize a
proton beam with a density comparable to the density of the hot background
electrons.
In view of this we will use $B_{\rm trap}$ as an estimate of the saturation
magnetic field strength.

The criteria (\ref{eq:trapcond}) and~(\ref{eq:MedvLoe}) predict the typical
amplitude of the magnetic field as one particular wave mode~$k$ saturates.
In a realistic situation the instability will involve a superposition of
wave modes and one should interpret $B$ as the amplitude that
follows from the power spectrum ${\cal I}_{\rm B}(k)$ of the field
fluctuations: $B^{2}/8 \pi\sim k {\cal I}_{\rm B}(k)$ with $k
\approx k_{\ast}$ for magnetization and $k \approx k^{\dagger}$ for
trapping.
The total magnetic energy in the unstable modes is
\begin{equation}
\label{eq:MagnU}
	U_{\rm B} = \frac{B^{2}}{8 \pi} = \int _{0}^{k_{\rm max}} {\rm d} k \: {\cal I}_{\rm B}(k)\,.
\end{equation}

\subsection{\label{sec:epsB}The equipartition parameter}

A measure of the strength of the magnetic field is the equipartition
parameter, which compares the energy density in the magnetic field with the
total energy density.

The protons dominate the energy budget, and the total available energy
density is
\begin{equation}
\label{eq:protonsh}
	e_{\rm p} = \int{\rm d}\vec{p}\,F (\vec{p})\gamma(\vec{p})m_{\rm p}c^2
		\simeq\gamma _{\rm b0}n_{\rm p}m_{\rm p}c^{2}\,,
\end{equation}
where the approximation is valid for $v_{z0}\ll v_{x0}$.

We define the proton equipartition parameter as
\begin{equation}
	\epsilon _B= \frac{B_{\rm trap}^{2}}{8 \pi e_{\rm p}}\,.
\end{equation}
Using Eq.~(\ref{eq:trapsat}) for the magnetic field, with
the definition~(\ref{eq:pfreq2}) for $\hat{\omega}_{\rm pp}$ and the approximation $v_{x0}\simeq c$
for relativistic shocks we get
\begin{equation}
\label{eq:eqpar}
 \epsilon _B= \frac{\hat{\omega}_{\rm pp}^{2}}{8 \Omega^{2}}\,.
\end{equation}
Then from Eqs.~(\ref{eq:plfreqratio}) and~(\ref{eq:Omapprox}) we have
\begin{equation}
\label{eq:epsB}
	\epsilon _B\simeq\frac{m_{\rm e}}{6 m_{\rm p}} \sim 10^{-4}\,.
\end{equation}

\section{\label{sec:discussion}Discussion}

The small value of the equipartition parameter~(\ref{eq:epsB}) implies that the
proton-driven Weibel instability in a background of relativistically hot
electrons saturates long before the magnetic field reaches equipartition
with the available proton free energy.
Equation~(\ref{eq:smallvzn2}) demonstrates that the proton-driven Weibel
instability in a hydrogen plasma is not a suitable candidate for the
mechanism responsible for `thermalization' of the incoming protons in the
shock layer.
In electron(-positron) beams in a hot pair background the instability
condition, $\alpha> 1$, allows a large beam velocity dispersion: $v_{z0}/v_{x0}\le 1$ if the number densities in the beam and in the hot background are of
similar magnitude \citep[see also][]{yoda87}.
Therefore, the Weibel instability is in principle capable of randomizing a
significant fraction of the beam momentum of an electron(-positron) beam, as
asserted in Sect.~\ref{sec:model}, whereas this is not true for a proton beam in
a hydrogen plasma.

In the limit of a cold, relativistic proton beam the properties of the
electrons and not those of the protons determine many of the results.
The electron plasma frequency determines the wave number~(\ref{eq:kbar}) of
the mode with the maximum magnetic field so that the electron skin depth
$c/\tilde{\omega}_{\rm pe}$ sets the length scale of the dominant mode.
This happens because the low inertia of the electrons makes them very
responsive to the perturbations of the protons.
The dispersion relation for the Weibel modes (Fig.~\ref{fig:sigma}), also
supports this view: the plateau around the maximum growth rate starts
roughly at a wave number $k \sim\tilde{\omega}_{\rm pe}/c$.
Studies that do not include the response of the background electrons
(by treating the protons as an isolated system) miss this point.
The peak magnetic field~(\ref{eq:trapsat}) does not contain any parameters
connected with the protons.
Therefore, proton beams in a hydrogen plasma generate nearly the same
magnetic field strength as electron(-positron) beams in an electron-positron
plasma (Fig.~\ref{fig:bsat}) {\em despite\/} the larger kinetic energy of the
protons.

\citet{gruz01} anticipated this when he excluded the case where a small
parameter in the theory might be important in his analysis of the Weibel
instability: our analysis shows that the relevant small parameter is
$\hat{\omega}_{\rm pp}^{2}/\Omega^{2} \sim 4m_{\rm e}/3m_{\rm p}$.
The result is a small equipartition parameter~(\ref{eq:epsB}).
In this respect our result is similar to the one found by
\citet[ p.~88]{sagd66}, who argued for the Weibel instability in a
non-relativistic plasma that the electrons have a quenching
effect on the ion instability.

In our analysis we have excluded electrostatic waves, which
could also play an important role in the shock transition zone
\citep{svbl02}.
In that case electrostatic Bremsstrahlung could be an alternative
explanation for the Gamma-ray Burst afterglow emission~\citep{schl03},
relaxing the need in synchrotron models for a high magnetic field strength.

\section{\label{sec:conclusions}Conclusions}

We have presented an analytical estimate of the magnetic field produced at
the end of the linear phase of the Weibel instability at the front of an
ultra-relativistic shock propagating into a cold hydrogen plasma, a
situation that applies to the external shocks that produce gamma-ray burst afterglows
in the fireball model~\citep{reme92}.
The magnetic field strength that we find is too weak to explain the observed
synchrotron radiation~\citep{grwa99}: the equipartition
parameter~(Eq.~\ref{eq:epsB}) is at least two orders of magnitude too
small.
This is radically different from the results for the Weibel instability in
pair plasmas.
The reason is that the contribution of the electrons to the electromagnetic
response of the plasma inhibits the instability of the protons.

The saturation magnetic field~(\ref{eq:trapsat}) which this low equipartition
parameter corresponds to is the magnetic field at the point where the
linear approximation breaks down and where non-linear trapping effects start
to limit the growth of the unstable Weibel mode.
After this happens, it is likely that the instability enters a nonlinear
phase or that another type of instability takes over: numerical
simulations~\citep{fhhn04} of similar plasmas show near-equipartition
magnetic fields behind the Weibel-unstable region in the shock transition.
The nonlinear phase would then be the dominant phase in electron-ion plasmas
and deserves further study to determine the physical mechanism and
the properties of the resulting magnetic field.

\begin{acknowledgements}
This research is supported by the Netherlands Research School for Astronomy
(NOVA).
\end{acknowledgements}

\bibliographystyle{aa}\bibliography{jw}

\appendix\section{\label{sec:asym}Asymmetric beams}

We consider asymmetric proton beams and show that we can neglect
the effect of the asymmetry for typical parameters.
We replace the proton distribution~(\ref{eq:distribution}) by
\begin{equation}
\label{eq:asymmetric}
\begin{split}
 F(\vec{p}) ={} &{\frac{n_{\rm p}}{2 p_{z0}}} \:
	\left[ \frac{1 + \Delta}{2} \: \delta(p_x-p_{x0}) + \frac{1 - \Delta}{2} \: \delta(p_x+ p_{x0}) \right]\\
	&{}\times\delta(p_y) \left[\Theta(p_z+ p_{z0}) - \Theta(p_z- p_{z0}) \right],
\end{split}
\end{equation}
with $0 \le\Delta\le 1$ the parameter measuring the asymmetry between
the two beams.
This asymmetry changes the dispersion relation of the waves to
\citep[{\it e.g.},][]{aaps75}:
\begin{equation}
\label{eq:asymmdisp}
	k^2 c^2 - \omega^{2} \left[ \: 1 + \chi _{xx}(\omega,\: k) -
	\frac{\chi _{xz}^{2}(\omega,\: k)}{1 + \chi _{zz}(\omega,\: k)}\: \right]
	= 0.
\end{equation}
The extra term $\propto\chi _{xz}^{2}$ appears in relation~(\ref{eq:asymmdisp})
because the waves are no longer transverse: the charge bunching of the
asymmetric beams produces a net charge density, leading to a component of
the wave electric field along the wave vector.

We will give the extra components of the susceptibility tensor and
publish a derivation elsewhere (Achterberg \& Wiersma, {\em in
preparation}).
The proton contribution to $\chi _{xx}(\omega, \: k)$ is the same as in the
symmetric case.
For a thermal electron background with an isotropic momentum distribution,
only the two proton beams contribute to the off-diagonal components of the
susceptibility tensor so that
\begin{equation}
\label{eq:offdiag}
	\chi _{xz}(\omega, k) = \chi _{xz, {\rm p}}(\omega, k) =
	\frac{\Delta\: \hat{\omega}_{\rm pp}^{2}}{\omega^{2} - k^{2} v_{z0}^{2}}
	\: \frac{k v_{x0}}{\omega}\,.
\end{equation}
This off-diagonal component vanishes in the symmetric case ($\Delta= 0$).
The longitudinal response of the beam-plasma system is contained in
\begin{equation}
\label{eq:longit}
	1 + \chi _{zz}(\omega,\: k) =
	1 - \frac{\tilde{\omega}_{\rm pe}^2}{\omega^{2}- k^{2} C_{\rm e}^{2}} - \frac{\hat{\omega}_{\rm pp}^{2}}
	{\omega^{2} - k^{2} v_{z0}^{2}}\,.
\end{equation}
Here $C_{\rm e}^{2} = 4 P_{\rm e}/3 n_{{\rm e}}m_{{\rm e}}h$ is the effective sound speed
in the hot electron gas.
In the ultra-relativistic case one has $C_{\rm e} \simeq c/\sqrt{3}$.
For proton beams in a hot electron background one has $\tilde{\omega}_{\rm pe}^{2} \gg\hat{\omega}_{\rm pp}^{2}$ and $|\omega| \le\hat{\omega}_{\rm pp}\ll k C_{\rm e}$ for the Weibel
instability (see Sect.~\ref{sec:growthrate}).
This implies that
\begin{equation}
 1 + \chi _{zz} \simeq 1 + \frac{\tilde{\omega}_{\rm pe}^{2}}{k^{2} C_{\rm e}^{2}}
	- \frac{\hat{\omega}_{\rm pp}^{2}}{\omega^{2} - k^{2} v_{z0}^{2}}.
\end{equation}

The resulting dispersion relation can
be written in terms of the dimensionless variable
\begin{equation}
\label{eq:Zdef}
	{\cal Z}(\omega, \: k) \equiv\frac{\omega^{2} - k^{2} v_{z0}^{2}}
	{\hat{\omega}_{\rm pp}^{2}}\,.
\end{equation}
One finds
\begin{equation}
\label{eq:asymdispZ}
	\left( {\cal Z} - {\cal Z}_{1} \right) \left( {\cal Z} + {\cal Z}_{2} \right) -
	\Delta^{2} \: {\cal Z}_{1} {\cal Z}_{2} =0.
\end{equation}
Here ${\cal Z}_{1,2}$ are defined by
\begin{equation}
\label{eq:Zdefs}
	{\cal Z}_{1} = \frac{k^{2} C_{\rm e}^{2}}{k^{2} C_{\rm e}^{2} + \tilde{\omega}_{\rm pe}^{2}}
	\; \; , \; \; {\cal Z}_{2} = \frac{k^{2} v_{x0}^{2}}{k^{2} c^{2} + \Omega^{2}} \end{equation}
with ${\cal Z}_{2} > {\cal Z}_{1}$ for the parameters considered here.
The solution of this quadratic equation,
\begin{equation}
\label{eq:Zsol}
	{\cal Z}_{\pm} = \frac{	{\cal Z}_{1} - {\cal Z}_{2}}{2} \pm\sqrt{\displaystyle\left( \frac{ {\cal Z}_{1} + {\cal Z}_{2}}{2} \right)^{2}
	- \Delta^{2} \: {\cal Z}_{1} {\cal Z}_{2}}\,,
\end{equation}
determines the frequency through
\begin{equation}
	\omega _{\pm}^{2} = \hat{\omega}_{\rm pp}^{2} \: {\cal Z}_{\pm} + k^{2} v_{z0}^{2}\,.
\end{equation}
The Weibel-unstable branch corresponds to the solution branch $\omega _{-}$ as ${\cal Z}_{-} < 0$.
In the symmetric case ($\Delta= 0$)
one has ${\cal Z}_{-} = - {\cal Z}_{2}$ which follows from~(\ref{eq:asymdispZ}),
and one recovers dispersion relation~(\ref{eq:approxdisp2}).
The stable branch $\omega _{+}$ is a modified (largely electrostatic) ion-acoustic wave.

\begin{figure}
\resizebox{\hsize}{!}{\includegraphics{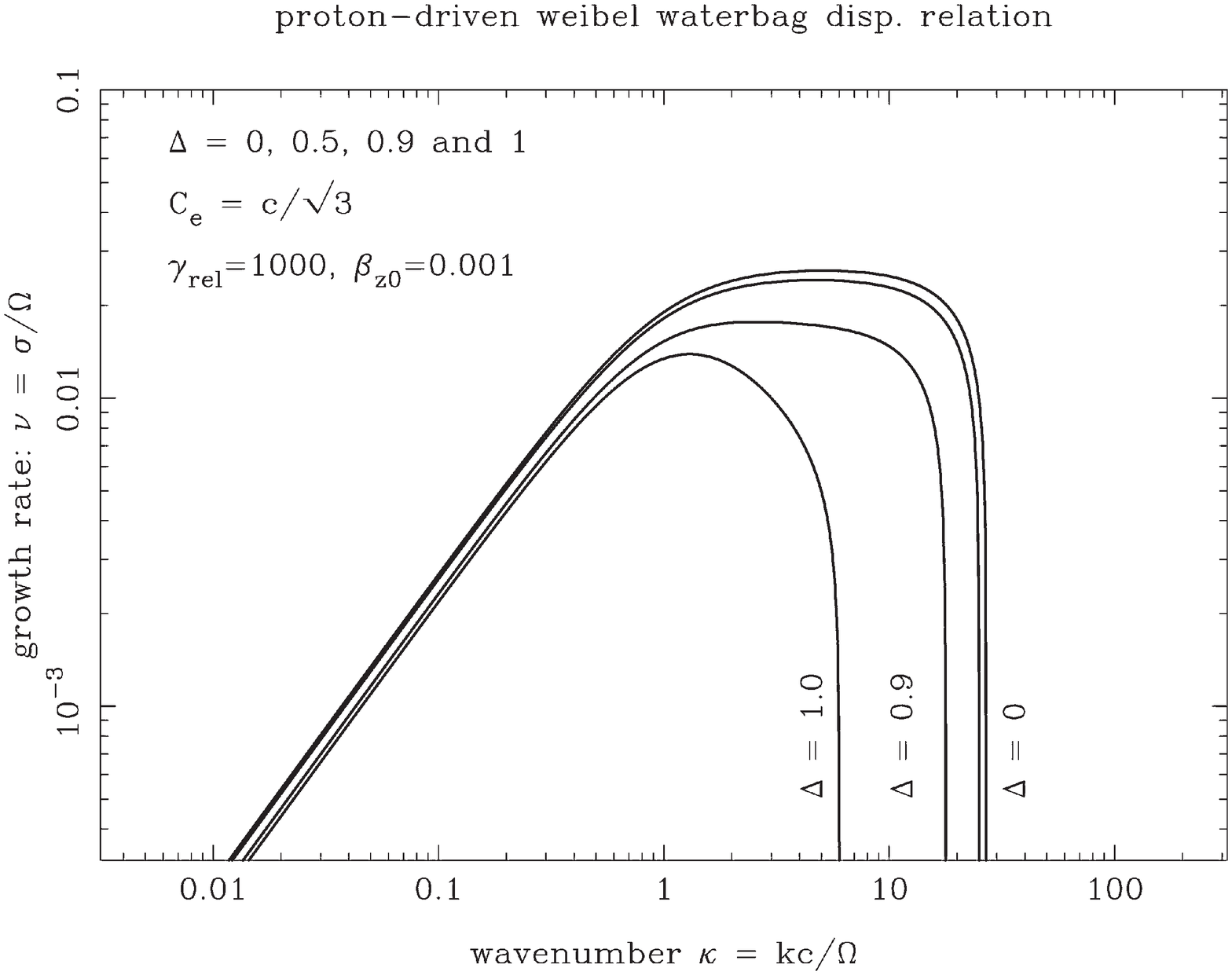}}
\caption{The growth rate as a function of wavenumber for
asymmetric proton beams in a relativistically hot electron background with sound
speed $C_{\rm e} = c/\sqrt{3}$.
Different lines correspond to different values of the parameter
$\Delta$, which measures the asymmetry between the beams.}
\label{fig:asym} \end{figure}

Although the asymmetry decreases the range of unstable wave numbers
(Fig.~\ref{fig:asym}) and lowers the growth rate with respect to the symmetric
case $\Delta= 0$, the change is small unless $\Delta\approx 1$: the case
where there is almost no reflection. For a single beam, $\Delta= 1$,
one has ${\cal Z}_{-} = {\cal Z}_{1} - {\cal Z}_{2}$, which
gives the following dispersion relation for $\nu^{2} = - \omega _{-}^{2}/\Omega^{2}$
in the ultra-relativistic limit with $C_{\rm e} = c/\sqrt{3}$ and $\Omega^{2} \approx\tilde{\omega}_{\rm pe}^{2}$:
\begin{equation}
\label{singleb}
	\nu^{2} = \frac{\hat{\omega}_{\rm pp}^{2}}{\Omega^{2}} \: \frac{ \displaystyle\kappa^{2} \: \left\{ \left( \frac{3 v_{x0}^{2}}{c^{2}} - 1 \right) -
	\kappa^{2} \: \left( 1 - \frac{v_{x0}^{2}}{c^{2}} \right) \right\}}{(3 + \kappa^{2})(1 + \kappa^{2})} - \kappa^{2} \frac{v_{z0}^{2}}{c^{2}}
	\,.	
\end{equation}
\end{document}